# MICROSTRUCTURE OF CONDUCTIVE CERAMICS $Al_2O_3.MnO_2.SiO_2$ IN VARIOUS CALCINATION TEMPERATURES

**Deski Beri[1], Rahmi Muthia[2], Ali Amran[1]\***

[a]Laboratory of Material Science, Department of Chemistry, Faculty of Mathematics and Science, Universitas Negeri Padang, Kampus UNP Air Tawar, Jl. Prof. Dr. Hamka, Padang, Sumatera Barat, Indonesia, 25131
Telp:+62-751-7058772
E-mail: deski.beri@gmail.com

**ABSTRACT**

Conductive ceramics played an important role in industries and manufactures due to their wide applications on electronics devices. Many synthesis approaches to prepare ceramics have been developing rapidly in the few decades. One of the simple and easy ways to prepare ceramics was using solgel method and varying calcinated temperatures. In this works we developed conductive ceramics, microstructure analysis by XRD and SEM, and conductivity measurement. Our important finding was an improvement in performance of microstructures, crystallinities and conductivities by increasing calcinated temperatures.

*Keywords: Calcinated temperature, Conductivity, Conductive Ceramics, Microstructure, Solgel*

## 1 INTRODUCTION

Electronics' industries were developing rapidly for decades due to the demands of electronics devices [1]. Annually, new products were launchings by many manufacturers, and this also followed by research and development for new products to come [2]. As part of electronics components, conductive ceramics played an important role to support the booms of electronics devices world. The demand of materials that could conduct electricity and signal at high speed with minimal produce of heats enhances research about conductive ceramics [3]. Conductive ceramics is useful to transmit signals and electricity, besides it could be used as electric charge's saver [4]. Therefore, conductive ceramics materials were used in electric wires, optical fibres, antennas, microchips, transmitters, semiconductors, resistors, capacitors and rechargeable batteries [5].

Many synthetic methods have been developed to produce conductive ceramics every year. One interested method was sol-gel method [6]. It has been used for decades to produce glass, aerosol, zeolite, composite, nanoparticle, and ceramics [7]. We used sol-gel method to produce conductive ceramics by making variation in temperature of calcinations [8]. Our aim was to design the best microstructures in terms of crystallinities, rheologies, and the relation to conductivity of conductive ceramics [9].

## 2 EXPERIMENTAL SECTION

### 2.1 Materials and Methods

In this research we used tetraethyl orthosilicate (TEOS) 99% Grade purchased from Sigma-Aldrich Co.Ltd., ethanol 98% pure, $Al(NO_3)_3 \cdot 9H_2O$, GR, and $Mn(NO_3)_2 \cdot 4H_2O$, GR, $HNO_3$ p.a, purchased from Merck.,GmbH., and Doubledistilled water purchased from Rafa Medika.

### 2.2 Sol-Gel preparation of ceramics $Al_2O_3.MnO_2.SiO_2$

TEOS 2.23 mL was added to the mixture of ethanol 1.3 mL and water 2.25 mL. Nitric acid 2N was added as a catalyst to complete the reaction:

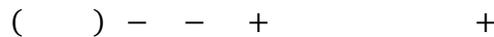

The mixtures were shaken for 1 hour to completely homogenize the miscible solution. 4.7 g $Al(NO_3)_3.9H_2O$ and 2.429 g $Mn(NO_3)_2.4H_2O$ were added to the solution and shaken continuously for 6 hour until the solution completely miscible. A homogenized clear sol was evaporate in $60^0C$ for 4 hour to become a wet gel, and then the wet gel was let to sit in room condition under glass box for 7-10 days to become xerogel.

### 2.3 Calcination process and characterization of ceramics $Al_2O_3.MnO_2.SiO_2$

Xerogel in above process was prepared 4 times to obtain 4 samples with similar treatment, then the 4 gels were calcined at 4 different temperatures. One gel was calcined at $900^0C$, one at $1000^0C$, one at $1100^0C$, and the last one at $1200^0C$, for 3 hours at their fixed temperature. The products were characterized using XRD and SEM.

### 2.4 Capacitance measurement of ceramics $Al_2O_3.MnO_2.SiO_2$

The calcined ceramics of $Al_2O_3.MnO_2.SiO_2$ were prepared in pellet form. The LCR meter was used to measure the capacitance value of ceramics.

## 3 RESULT AND DISCUSSION

### 3.1 Synthesis Result

The sol prepared by sol-gel method was clear, translucent and transparent as shown on Figure 1, and the average time needed to form xerogel was 8 days. The physical appearance of the sol was so homogeny as well as true solutions. At hydrolysis process, the reactions is believed to occur as follow

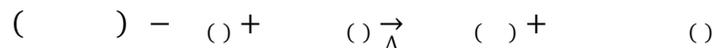

This reactions was followed by the reaction with alumina and manganese salts

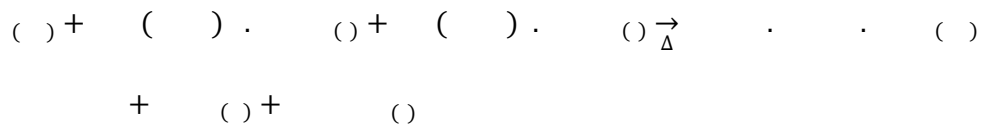

This process was called colloids methods due to the process used nitric salts as the precursor either alcoxides ion. Sol of . . ( ) then being evaporated in water bath at 60°C for 4 hour to help condensation process, and to let the shrinkages process taking place. After drying process for about 7-10 days, the clear translucent xerogel was losen from the tubes.

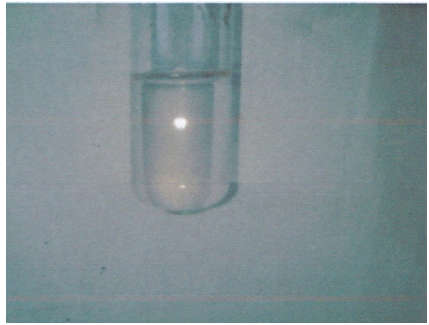

**Figure 1 Translucent sol of act ceramics . . ( ) prepared by sol-gel methods**

### 3.2 Calcination and Characterization Process

Powder xerogels, after grinded in mortar, then calcined in the furnace. Calcination process were done for four samples in four calcination temperatures: 900°C, 1000°C, 1100°C and 1200°C, respectively. Each sample was kept in fixed temperatures for 4 hours. After calcinations process, the appearance of ceramics 900°C, 1000°C and 1100°C were all similar with black coloration in bulky form. However, the appearance of ceramics calcined at temperature of 1200°C was different, it formed a big bulky form after melted. The appearance picture is presented in Figure 2.

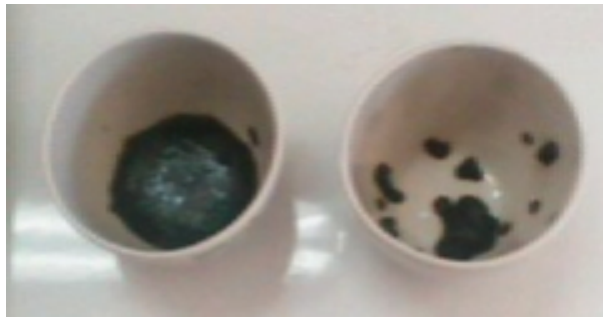

**Figure 2 The appearance of ceramics . . ( ) after the calcination process. (left) black bulky form after calcined at temperatures of 900°C, the similar form was also appeared for ceramics after calcined at temperature of 1000°C, and 1100°C. (right) big bulky melted form of ceramics after calcined at 1200°C**

Under calcination temperature of 900°C, 1000°C and 1100°C, the ceramics formed the compact structures, since the visual appearance of the materials looked more homogeny. But under calcination temperature of 1200°C, the melting process broke the structures and led the metals to react with the excess of oxygen in the furnace, and the oxidation process took place in the following formula:

$$\cdot\quad\cdot\quad(\ )+\ \xrightarrow{\Delta}\ (\ )+\ (\ )+\ (\ )$$

Compounds on the left side were in compact structures, whereas compounds on the right side were in isotropic oxide form. The detailed experiments about the transition structures were presented in XRD data at Figure 3. Starting from the lowest temperature (900°C), the peaks were showed at 2θ of 26°, 34°, 56°, and 66°. Increasing calcination temperatures made two peaks at 37° and 42° appeared. However, at calcination temperature of 1200°C, all peaks were collapse to form isotropic liquids called amorphous.

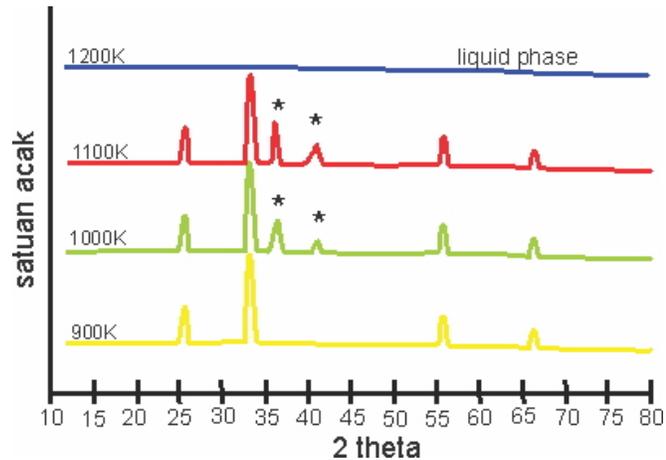

**Figure 2 XRD spectrum of ceramics . . ( ) under various calcination temperatures**

Using PCPDFWIN, we analyzed the XRD spectral and we found that ceramics formed the cubic garnet structure. Double twin peaks that appear in temperature of 1000 and 1100°C were responsible for the reposition of Al and Mn in the central and in the cubic corners. Our result were suited to compared with Sawada [10]. Reference to ICDS data No 50621, our product was found to be cubical garnet as presented in Figure 4. Highly graded crystalline structures were gradually become better when calcination temperature increased. The best crystalline structure was obtained at calcination temperature of 1100°C. At this temperature, crystalline structure was at the highest performance before the structure was collapsed to a frustratic isotropic phase, and melting at calcination temperature of 1200°C.

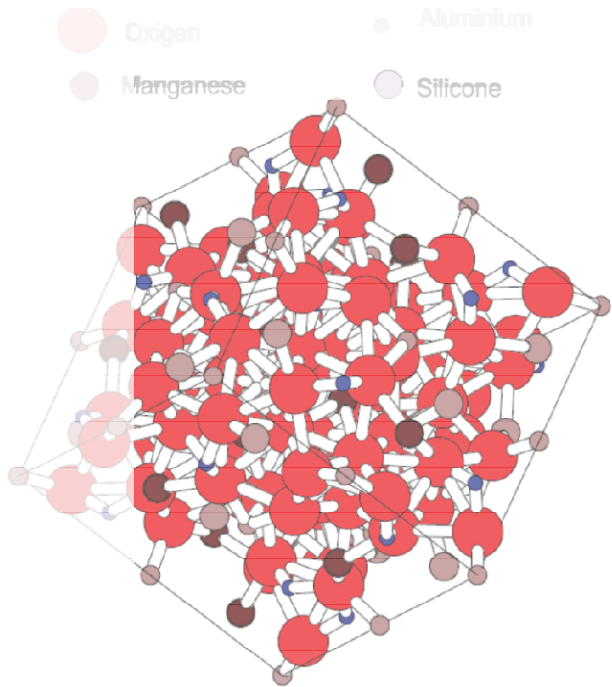

**Figure 3 Structure of ceramics . . ( ) after being referenced to PCPDFWIN and NIST data**

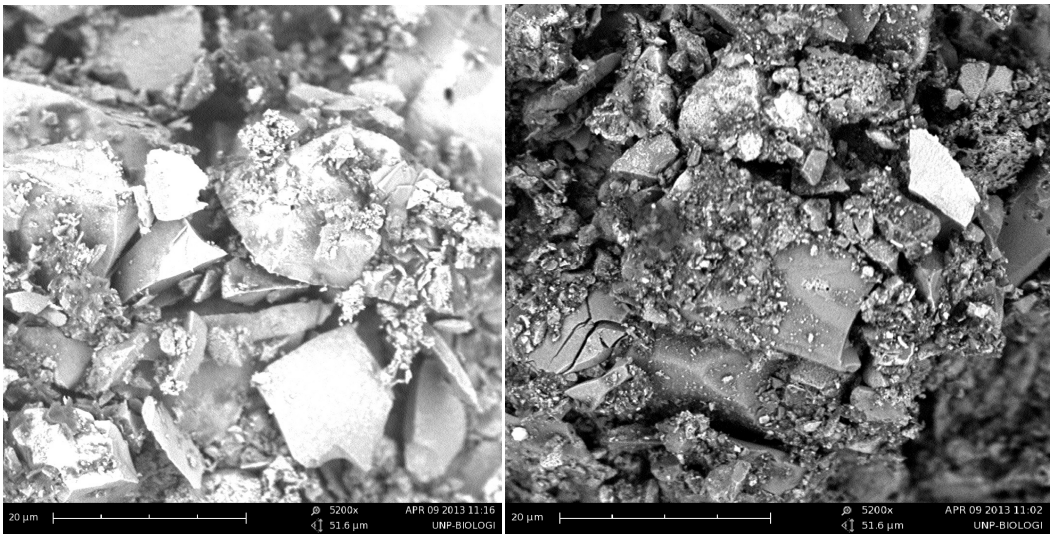

(a) (b)

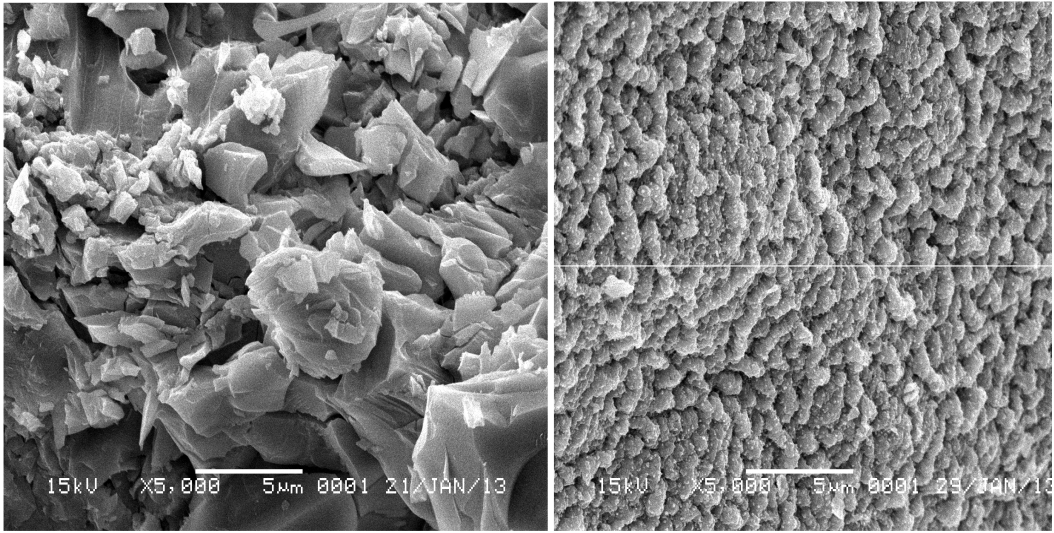

**Figure 4 SEM picture of ceramics . . ( ) under various calcination temperatures (a) 900⁰C, (b) 1000⁰C, (c)1100⁰C and (d) 1200⁰C**

SEM pictures show that ceramics under calcination temperature of 1100⁰C, was microscopically homogeny, as could be seen in spreadly dispersed homogeneous bulky particles. Ceramic crystals were growth in two steps, those were; initiation step by nucleation process, and then followed by crystal growing in-radially around the nuclei. From SEM picture it could be seen that the optimum calcination temperature to form crystalline phase ceramic of . . ( ) was at 1100⁰C. Complete figure of SEM data presented in Figure 5.

**3.3 Capacitance Measurement of Ceramics . . ( )**

Capacitance value of ceramics . . ( ) was measured using LCR meter. Measurement conducted by input voltage of 2 volt with time variation of 60 seconds. Capacitance value vs calcination temperature was plotted as curve in **Figure 6.**

Capacitance measurement show that the maximum capacitance was at calcination temperature of 1100⁰C. This could be explained that at 1100⁰C, crystalline structure of ceramic was at good crystallinity, so this made the electron transfer from metal to ion in compact structure become much better. However at calcination temperature of 1200⁰C, structures collapsed to become amorphous structure, and it caused the previous conductive ceramic become an isolator.

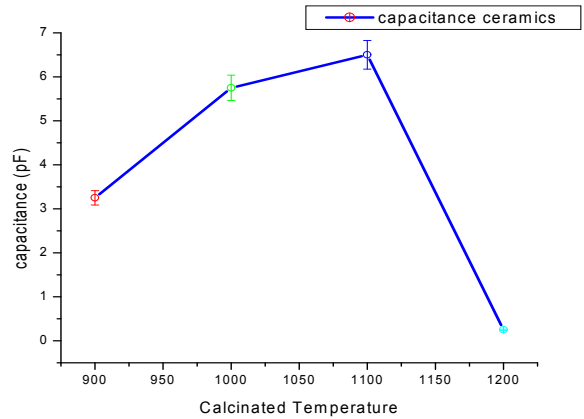

**Figure 6** The relation between capacitance of ceramics . . ( ) and calcination temperature. (maximum capacitance at calcination temperature of 1100$^0$C)

## 4  CONCLUSIONS

In this research we have successfully prepared conductive ceramics by sol-gel methods. Sol was prepared by hydrolysis precursor TEOS, then alumina and manganese salts was added and mixed homogeneusly. Sol was evaporated at 60$^0$C to formed xerogel. The appearing of sol and xerogel was clear, translucent and transparent. Xerogels were prepared four kinds in similar procedure and each were calcination in 900$^0$C, 1000$^0$C, 1100$^0$C and 1200$^0$C to form black bulky materials. XRD measurement presented that the grade crystallinity were improved gradually from 900$^0$C to 1100$^0$C, but at 1200$^0$C the structure was collapsed to form frustrated isotropic phase and become amorphous, this analysis was supported by SEM data. Capacitance measurements by LCR meter also confirmed our result that the value of capacitance was increased by increasing calcination temperature, but at 1200$^0$C, at amorphous phase, the capacitance value drastically reduced due to low graded crystallinity.